\documentclass[12pt,a4paper,aps]{article}
\usepackage{a4}
\usepackage{epsfig}
\usepackage[hmargin=2cm,vmargin=2cm,nohead]{geometry}

\usepackage{graphicx}
\usepackage{amssymb}
\usepackage{epstopdf}
\newif\ifpdf
  \ifx\pdfoutput\undefined
  \pdffalse
\else
  \pdfoutput=1
  \pdftrue
\fi

\ifpdf
  \DeclareGraphicsExtensions{.pdf,.jpg,.JPG,.jpeg,.gif}
\else
  \DeclareGraphicsExtensions{.ps,.eps,.eps.gz,.ps.gz}
\fi

\ifpdf
\usepackage[colorlinks=true, pdfstartview=FitV, linkcolor=blue, 
            citecolor=blue, urlcolor=blue]{hyperref}
\hypersetup{pdfstartview={Fit -32768},
  pdfpagemode=UseOutlines,
  plainpages=true,
  bookmarksnumbered=true,
  bookmarksopen=false
  }

\begin{document}

\title{Development Towards Sustainability: \\How to judge past and proposed policies?} 

\author{
Michael Dittmar\thanks{e-mail:Michael.Dittmar@cern.ch},\\
Institute of Particle Physics,\\ 
ETH, 8093 Zurich, Switzerland\\
\date{August 30, 2013} \\
Submitted to the journal: Science of the Total Environment
}
\maketitle

\begin{abstract}
\noindent
Most countries have, at least since the 1992 United Nations summit in RIO, adopted some vague ``sustainable development" policies. The goals of such policies are to combine 
economic growth with social development, while  
protecting our fragile planetary life support system. 
 
\noindent
The scientific data about the state of our planet,  presented at the 2012 (Rio+20) summit, documented that 
today's human family lives even less sustainably than it did in 1992. 
The data indicate furthermore that the environmental impacts from our current economic activities are so large, that we are approaching situations where potentially controllable regional problems  
can easily lead to uncontrollable global disasters. 
 
Despite these obvious failures, our political global leaders and their institutions are continuing the same 
``sustainable development" policies, which are now supplemented by equally vague ideas about 
future ``green economies".

\noindent
Assuming that (1) the majority of the human family, once adequately informed, wants to achieve a ``sustainable way of life" and (2) that the ``development towards sustainability" roadmap will be based on 
scientific principles,  one must begin with unambiguous and quantifiable definitions of these goals.  
As will be demonstrated, the well known scientific method to define abstract and complex issues by their negation, satisfies these requirements. 
Following this new approach, it also becomes possible to decide if proposed and actual policies changes will make our way of life less unsustainable, and thus move us potentially into the direction of sustainability.  Furthermore, if potentially dangerous tipping points are to be avoided, 
the transition roadmap must include some minimal speed requirements. 
Combining the negation method and the time evolution of that remaining natural capital in different 
domains, the transition speed  for a ``development towards sustainability"
can be quantified at local, regional and global scales. 

The presented ideas allow us to measure the rate of natural capital depletion and the rate of restoration 
that will be required if humanity is to avoid reaching a sustainable 
future by a collapse transition.  
Unfortunately, the existence of quantifiable methods and tools in no way guarantees that 
they will be used in changing the direction of our journey. 
\\

\end{abstract}


\newpage
\section{Introduction}
During the 1960s, while the disastrous impacts from colonialism, the First - and Second World Wars, and 
the dangers of total nuclear annihilation during the cold war period were still in the minds of the 
people, a new threat for the future of life on our planet emerged. 

A rapid exploitation and depletion of the natural capital resulted in global air pollution problems,
reduced the availability of clean water even in wealthier areas, and caused 
many people to begin doubting the system of industrial agriculture, which increased quantity but at the same time ignored the issue of food quality and contributed to health issues like obesity.
 
Those threats were understood by many natural scientists, mostly working in the richer countries, who realized that the globalized human impact was reaching proportions well beyond the repair capacity of our planetary life support system. Many recognized that the negative and unintended side effects of our human technological success story became a threat for the biosphere itself and thus for our own survival. As a result, a new and global civil society movement emerged. People started to point their fingers at the dark sides of a growth-based world economy with the resulting blind consumerism and the related disastrous effects from our industrial agricultural system.  

Scientific reports demonstrated that continued human population growth, combined with increased human affluence, was in general and in the long term outpacing technological efficiency improvements. This was quantified with the IPAT equation introduced by Ehrlich and Holdren in 1971,~\cite{Ehrlich1971}. They proposed to quantify the overall human impact accordingly to I(impact)=P(population) x A (affluence) x T(technology). The affluence (multiplied by technology) tries to quantify the average person impact on certain renewable and non renewable resources, our natural capital. Taking the oil consumption 
as one item of our affluence, one finds that the oil usage of an average human being in 2012 
corresponds to an oil use of about 2 liter per day. The corresponding daily oil consumptions for some high, medium and low affluence countries demonstrate huge difference in the per capita impact 
in rich and poor countries: USA (9.3 l), Switzerland (4.8 l), Italy (3.5 l), Poland (2.3 l), China (1.3 l), India (0.5 l), Nigeria (0.35 l) and Niger (0.01),~\cite{Oilpercapita}.  
While neglecting the relation of this impact to the remaining natural capital and its quantitative depletion, the obvious consequence of this equation was that a stable and sustainable system requires an overall human impact smaller than the regional and global natural growth yield. 

The first computer based ``world models",~\cite{Forester}, appeared at around the same time. The researchers tried to investigate the effects of the limited, but quantitatively badly known, planetary resources and the pollution consequences of their 
uses for the overall functioning of our global and growth based economic system. 
The results, published 1972 in the book ``The Limits of Growth", \cite{Meadows1972}, indicated that our economic growth-based system will reach, either by resource depletion or by the resulting environmental pollution, its unavoidable limits of growth during the first half of the 21st century, 
followed by a painfully dramatic decline.
 
Catton, \cite{Catton1980}, and others concentrated their research efforts on the relation between the regional and global biological carrying capacity for different species, including humans, and the unsustainable use of the natural capital. Catton concluded that an ``overshoot" (overuse) 
of the local carrying capacity is unsustainable and can only be maintained as long as imports, which usually exploit and deplete the natural capital elsewhere, can continue.
For example, the introduction of the oil-based cheap and fast globalized transport system increased the 
ability to import food and other resources over large distances by a significant factor. As a result, 
we made our ``way of life" dependent on an unsustainable transport system and came 
to the point that we are now confronted with a much larger overshoot problem -- a global 
overshoot problem.

These early scientific warnings about the dangers of our unsustainable path were widely discussed 
by many people in our society. However, probably also due to the very abstract theoretical 
basis and due to the absence of quantitative calculations, the scientific basis for these warnings were dismissed by most economists and politicians. While essentially nobody dared to claim that exponential growth on a finite planet can continue forever, the absence of accurate quantitative data about resource limits in the 
1970s allowed the opponents to claim that undiscovered (energy) resources, while certainly limited, are 
still so large that theoretical growth limits are still hundreds of years away~\cite{Simon}.

Similar views were accomplished with respect to the problems related to the 
increasing environmental degradation, the damages to the biodiversity and the 
toxic waste and pesticide pollution resulting from the industrial agriculture.
The potentially very negative impacts were acknowledged to exist, 
but the resulting mostly local problems were considered, even up to this day, to be minor when  
compared to the economic benefits and to the size of the overall remaining natural capital. 
And ignoring the problems received further support when the 
related potential dangers for human health were associated with large uncertainties, and when
some technological advances helped to reduce at least some ÒvisibleÓ pollutions effects.

The growing evidence in the 1980s for (1) global climate change due to ever-increasing CO$_{2}$ emissions, (2) dramatic biodiversity losses and (3) the problems related to our unsustainable industrial agricultural system, resulted in policies favoring programs of ``sustainable development".
Under the leadership of the United Nations and some governments, ``sustainable development" programs became widespread and culminated into the UN conference on ``sustainable development" 
in 1992 in Rio de Janeiro~\cite{Rio1992a}. During this conference essentially all political world leaders declared the preservation of the natural environment and for future generations as their goal~\cite{Rio1992b}.
Due to the vague nature of these 
goals however, the policies did not result in a utilizable roadmap towards a sustainable 
way of life, nor did governments in any way accept responsibility for achieving the stated 
goals.\\

Predictably, the 2012 UN RIO+20 follow-up conference,~\cite{Rio2012}, was faced with an abundance 
of devastating scientific evidence regarding the failure to move toward the stated goals (see 
section 2 for some examples),~\cite{UNEP2012a}. These data demonstrated the total failure of the global policy makers to protect our planetary life support system.  The absence of essentially all political world leaders at the 2012 summit indicated further that not even the spirit from the 1992 conference survived\footnote{As expressed by the  
Guardian(~\cite{GuardianRio}): ``After Rio, we know. Governments have given up on the planetÓ.}. 
Unfortunately this disinterest has now spread also to the scientific community.  In contrast  
to the strong 1992 declaration ``Warning to humanity",~\cite{Warning}, signed by hundreds of Noble laureats and supported by more than 1700 world renown scientists,  the 2012 follow-up led to no statement 
of a common view within the scientific community demonstrating an even smaller acceptance of responsibility to inform the world population and pressure governments to act.  

The current situation within the scientific community is especially disillusioning, 
as, compared to 1992,  precise data from all around the planet demonstrate the unfolding disaster.
Despite these failures, however, the civil society and many scientists have still not given
up, and more realistic ideas about the best paths towards sustainability are emerging and 
developing,~\cite{civilsociety}. 

In the following sections, after presenting some facts about the disastrous results from 
``sustainable development" policies, we will introduce {\bf the negation concept}, \cite{negation}, which allows us to define {\bf ``sustainability"} and {\bf ``development towards sustainability"} in  
an unambiguous and quantifiable scientifically useful way. 
{\bf This new approach}, as we will demonstrate, allows us not only to quantify potential and real achievements of proposed policies, but gives us also a basis that allows us to determine 
the time frame within which a successful transition to sustainability must be achieved --the alternative 
being an extremely painful transition to sustainability imposed on us by Òmother natureÓ.
      
\section{The failures of ``sustainable development" policies.}

Policies claiming to follow ``sustainable development" principles emerged during the 1980s. Such policies  
became part of the official United Nations program with the 1987 Brundtland Report, \cite{Brundtland} and the 1992 
RIO earth summit, \cite{Rio1992a}, where one finds the following definition: \\
{\it ``Sustainable development is development that meets the needs of the present without compromising the ability of future generations to meet their own needs."}. \\
Since the aim of this paper is to present an unambiguous and quantifiable definition of sustainability, we refer to the literature regarding different interpretations of  ``sustainable development" 
policies~\cite{Brundtland}. However, it appears to be useful to start this paper with some facts about the devastating results from the 20 and more years of ``Sustainable development" policies. More details can be 
obtained from the detailed 2012 UNEP report, \cite{UNEP2012a}.  The following data are presented in such a way that they can be directly associated with the overall human impact, as quantified by the I=PAT equation, \cite{Ehrlich1971}: 

\begin{itemize} 
\item The world population (P) has increased from 3.9 billion people (1971) to 5.5 billion 
(1992) to about 7 billion people at the time of the RIO+20 conference in 2012. According to UN demographers, and without any catastrophes, another 1-1.5 billion people will be added during the next 20 years, \cite{UNpopcensus}. 
\item The annual world energy consumption of non renewable energy resources, a first order approximation of the global human impact, has almost doubled between 1972 and 2012. This increase 
happened in roughly equal proportions during the two 20 year periods, \cite{IEA2012}. 
\item The CO$_{2}$ level in the atmosphere has increased by roughly 1.5 ppm per year from 330 ppm (1972) to 359 ppm (1992) and by 1.9 ppm per year to 397 ppm (2012). 
Without revolutionary changes in our economic system, or a lasting dramatic world economic crisis,  
the international organizations predict that the CO$_{2}$ concentration will continue to increase by similar annual amounts during the next decades, \cite{IPCC2012}.  
\item The total amounts of slow-to-decompose solid waste, 
highly toxic chemical and radioactive nuclear energy waste and other unwanted side products of our industrial way of life appears to be less well documented. However our garbage ``monuments", like the new ``continent" made of plastic waste in the center of the pacific ocean, \cite{plasticwaste}, 
constructed after 1992, demonstrate that the approaches to solve our waste problem failed. 
According to the predictions from the World Bank, our global waste problem will continue to increase during the next decades, commensurate with the growth of the global economy, \cite{Worldbank2012}. 
\item The remaining pristine forests on our planet continue to be eliminated at dramatic rates. 
The only relief seems to be in regions where essentially all of the old and economically most 
valuable trees have already been removed,~\cite{UNEP2012b}.
\item 
The ocean fish stocks have continued to decline rapidly and 85\% of the world's fish stocks are
``over-exploited, depleted, fully exploited or in recovery from exploitation",~\cite{BBCfish}. 
In addition, land erosion, desertification, and river water pollution, with the resulting dead zones near the river deltas, have increased dramatically, \cite{UNEP2012c}.  
\item Biodiversity loss has now been transformed into mass extinctions, \cite{UNEP2012d},
and the latest measurements  from the European Environmental agency (July 2013) show that butterfly populations in 
Europe have declined by about 50\% during the last 20 years, \cite{butterflies}.  
\end{itemize}

This list shows that, whatever the original goals were, the world's ``sustainable 
development" policies have barely reduced the rate of acceleration of the suicidal 
destruction of our life support system -- if they have even done that. The obvious failure 
of these policies and the resulting crisis are well summarized within the 
UNEP report,~\cite{UNEP2012a} as follows:

{\it ``As human pressures within the Earth System increase, several critical thresholds are approaching or have been exceeded, beyond which abrupt and non-linear changes to the life-support functions of the planet could occur. This has significant implications for human well-being now and in the future."}

\section{Defining ``sustainability" and ``development towards sustainability"} 

The following ideas are based on the assumption (1) that a large majority of our global human family, once ``adequately informed", would reject proceeding any further down the current 
path towards extinction, (2) that the ``information" involved must be based on well-established
science, and (3) that ÒsustainabilityÓ and Òdevelopment towards sustainabilityÓ must be defined 
in an unambiguous and quantiÞable way.\\

As will be demonstrated, such definitions can be obtained with the help of the well founded scientific method to {\bf define an abstract concept by its negation}, \cite{negation}, and the fact that our way of life is either ``sustainable" or ``unsustainable".

Thus, instead of trying to define what we mean by sustainable living, we simplify our task by describing unsustainable aspects in our way of life. This negation approach allows to investigate and quantify the unsustainable problem at the individual, local, regional and global level. Here we are focusing 
four important\footnote{One can certainly add areas where the natural capital is used in unsustainable 
ways.} and easy recognizable unsustainable key areas in our industrial way of life: 
\begin{itemize}
\item The usage of ``non renewable" and finite energy resources (oil, gas, coal, uranium), and 
the usage of non-renewable mineral resources as long as the recycling of those resources 
is based on non-renewable energy resources.
\item The unsustainable use of energy from what seem renewable sources like water, wood, wind, sun and soil. The past and present use of timber, resulting in the destruction of pristine forests, 
is a well documented example of such unsustainable practices. 
\item The use of industrial agriculture, unimaginable without the direct and indirect use of fossil fuels, 
damaging to soil, lakes, rivers, and oceans and threatening to what remains of the planet's already fragile biodiversity.
This failure is demonstrated by soil erosion maps and dead ocean zones near the river 
deltas,~\cite{riverdeltas}. 
It is a sad point to note that this system was developed, practiced and distributed thanks to the research and the direct help from the majority of our related 
scientific institutions from within the so called most advanced rich nations.   
\item The production of non-compostables and very-slow-to-decompose toxic waste which is 
the result of misusing and depleting non-renewable natural capital. 
Well known examples are products like asbestos, plastics, chemical and radioactive toxic waste and, 
with the current level of emissions, greenhouse gases like CO$_{2}$.   
\end{itemize} 

With the knowledge that the stability of systems depends on their weakest elements, one can conclude that our way of life remains unsustainable as long as the activities in any of the above key areas continue. It appears that the ``negation" approach not only provides us with a unique 
definition of sustainability, but that it presents us directly with the requirements to find a possible roadmap 
to sustainability.  Such a roadmap must indicate the directions for all key areas and must demonstrate 
how our unsustainable way of life will become less and less unsustainable.

However, as ``quantum" transitions are only relevant on microscopic scales, our efforts 
to change directions are confronted with large moments of inertia. Nevertheless, and with respect to the different key areas, one knows that the time left for changes is not infinite. Thus,  a potentially successful roadmap must include some minimal transition speed requirements. 
Some ideas on how the minimal speed for a successful ``development towards sustainability" can be constrained will be presented in the next section. 
 
\section{Speed requirements for the successful roadmap:\\
``development towards sustainability"}
 
  

%

After having demonstrated that the negation principle enables 
us to define sustainability and the roadmap from our current unsustainable system 
into a sustainable system in an unambiguous way,  we are now presenting some principle ideas on how the minimal transition speed for a rational and successful ``development towards sustainability" can be determined\footnote{
The word ``development" is used to describe the transition from one system into another one.}. \\

The problems we are facing can be visualized easily with the simple example of a sailing trip to a distant island:\\ 
The participants of the trip consider the boat with some of its elements 
as their natural capital, existing in different and only partially exchangeable ``currencies". 
The simplified list contains a finite amount of fresh drinking water, food, some tools for 
fishing, maintenance and repair, some solar panels, and some diesel oil for a backup electric energy generator, for cooking and heating and perhaps also for the ship engine in times of little wind and during some emergency situations.  Some of the elements, like drinking water and food from fishing, are ``renewable" at a certain and fluctuating rate. Other resources like diesel oil are finite and non renewable. 
Assuming that neither the distance nor the travel speed are known precisely, most crew members might want to have an idea about their daily impact on the existing reserves and in comparison to their replacement rate on board and the expected length of the trip.
Perhaps they also want to participate, within a democratic decision making process, on the resource use during the trip. Depending on the depletion rate of the different resources, separated into different unexchangeable items, they can make a rationally motivated decision about the continuation of the trip, 
resulting in a possible resource exhaustion, or decide to turn back safely to their harbor. Despite warnings expressed by older crew members about uncalculable events, like unfavorable wind directions, which might require some extra spare capacity, it is decided that one should turn back home when the natural capital is reduced to about 50\% of its original value. \\

The above example leads us directly to the procedure on how the transition speed, for our journey into sustainability, can be determined from the measurable remaining natural capital 
and for each of the different unsustainable key areas.

We just need to know the actual locally and globally accessible resource data, and this in relation 
to their actual local and regional use. Furthermore, we need to know how some non-renewable resources can potentially be replaced with others. It is obvious that the calculable speed requirements depend on the accuracy of the
available resource data and the associated uncertainties. Depending on what is considered 
as an acceptable risk behavior, remaining uncertainties should require higher transition 
speeds, than would be estimated from precise data. 

Starting with the negation approach, we could define different key areas with potentially little overlap. 
As a first step, we need to separate the existing natural capital into its renewable and non-renewable 
components following formula (1), and their possible sustainable and unsustainable yields.     
\begin{center}
\begin{equation}
Natural~ capital(time) = non-renewable~capital (time) + renewable~capital(time)
\end{equation} 
\end{center}
The non-renewable sector gives zero interest rate and can be used and depleted at particular 
rates, which usually depend on the available technology and the state of the remaining capital.
  
Concerning the renewable sector, like for example the usage of timber, one finds that 
the annual forest growth rate provides us roughly with the actual usable timber yield. 
Assuming that we make use of a well established forest, with a limited size, the sustainable 
usage, and ignoring safety factors, is constrained by the annual growth rate multiplied with some efficiency factors. \\

When looking at the situation of the remaining energy resources, one needs to 
complicate formula (1) by some often unknown weight factors. These weight factors 
depend on the possibility to replace for example transport oil, with some alternatives like 
alcohol made by biological processes, or to transform heating oil into transport oil, while the 
heating system will be based on burning of timber. 

Next we need to relate the annual resource usage to the potential annual resource yield, 
which in general is a function of the actual existing natural capital and its extraction technology.  

Once the resource data, including their uncertainties, have been determined, the remaining 
global resource lifetime can be obtained easily as indicated in Figure 1. 

\begin{figure}[h]
\begin{center}
\includegraphics[width=15cm]{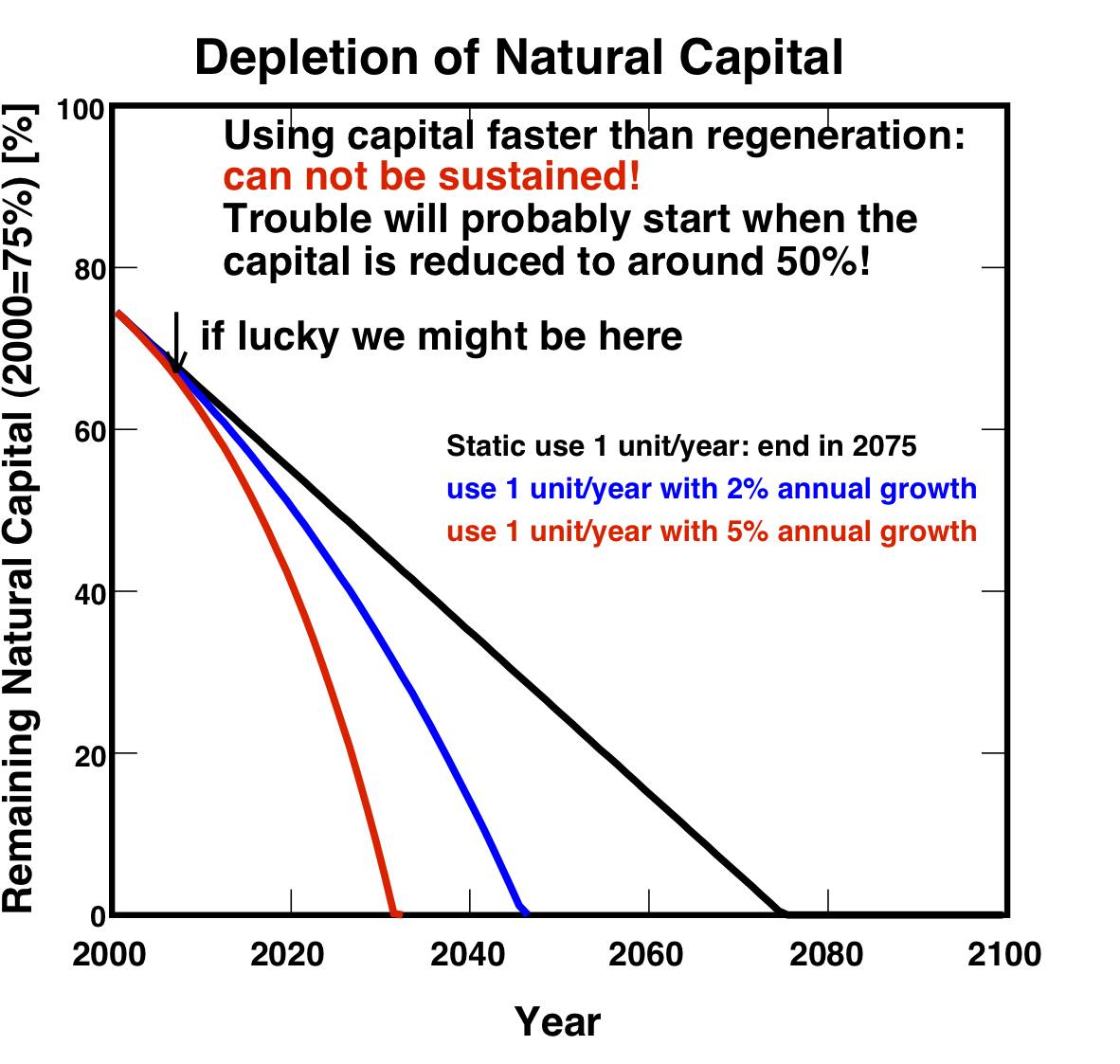}
\caption{Potential time dependent resource depletion curves assuming that the original capital had been reduced in the year 2000 already to about 75\% of its original size.  
The straight line corresponds to a constant annual depletion rate of 1 unit per year. In comparison 
the lifetime can be reduced significantly when unsustainable practices are allowed to grow 
exponentially with a 2\% and 5\% annual growth rate respectively.}
\end{center}
\end{figure}

When analyzing the current unsustainable energy sector, one might consider separating 
the different services provided for the current regional and global needs, 
and to divide it into the different renewable and non renewable resources.    

For example one could separate the mobility services, the heating services and  
the needs satisfied with electric energy. One also needs to investigate how the  
different energy resources are used for the different industrial and private applications. Globally one finds 
that more than 80\% of the energy services are unsustainable as they are based on non renewable resources. 
When looking at the different types of energy uses, one finds that especially the transport sector is currently 
almost 100\% based on the usage of oil. 

Alternatively, one can look into the actual energy sector in different countries and how renewable 
and non renewable energy resources are satisfying the actual annual usage.
Taking Switzerland as an example, \cite{SwissE}, one finds that oil provides about 55\% of the energy needs. Gas and 
uranium contribute about 14\% and 10\% respectively. The remaining 20\% are shared between 
the renewable sectors with 15\% originating from hydropower and 5\% from wood energy. 
If one assumes that all the energy services provided currently from oil, and ignoring transformation losses, would have to be supplied from biomass and renewables, other than hydropower, one finds 
that the timber usage would have to be increased for example by more than a factor of 10. Such a demand for timber would destroy the forest cover of Switzerland within a few years, and would result in subsequent unimaginable environmental disasters. In comparison, to such totally impossible transitions,  
current estimates indicate that a careful managed forestry could perhaps increase 
the sustainable use of timber by about 20\%, corresponding 
to a relative increase in the current energy mix from 5\% to about 6\%. 
We are thus confronted with a realistic Swiss energy impact (or energy footprint) which is at least 10 times larger than the energy imbedded in the existing entire biological timber growth rate within Switzerland. 
This simple calculation demonstrates that current estimates of Switzerland's ecological footprint, \cite{FootprintCH}, which claim an overshoot of about four, are dramatically wrong as they underestimate the actual overshoot related only the oil energy use by at least a factor of 10! 

\begin{figure}[h]
\begin{center}
\includegraphics[width=15cm]{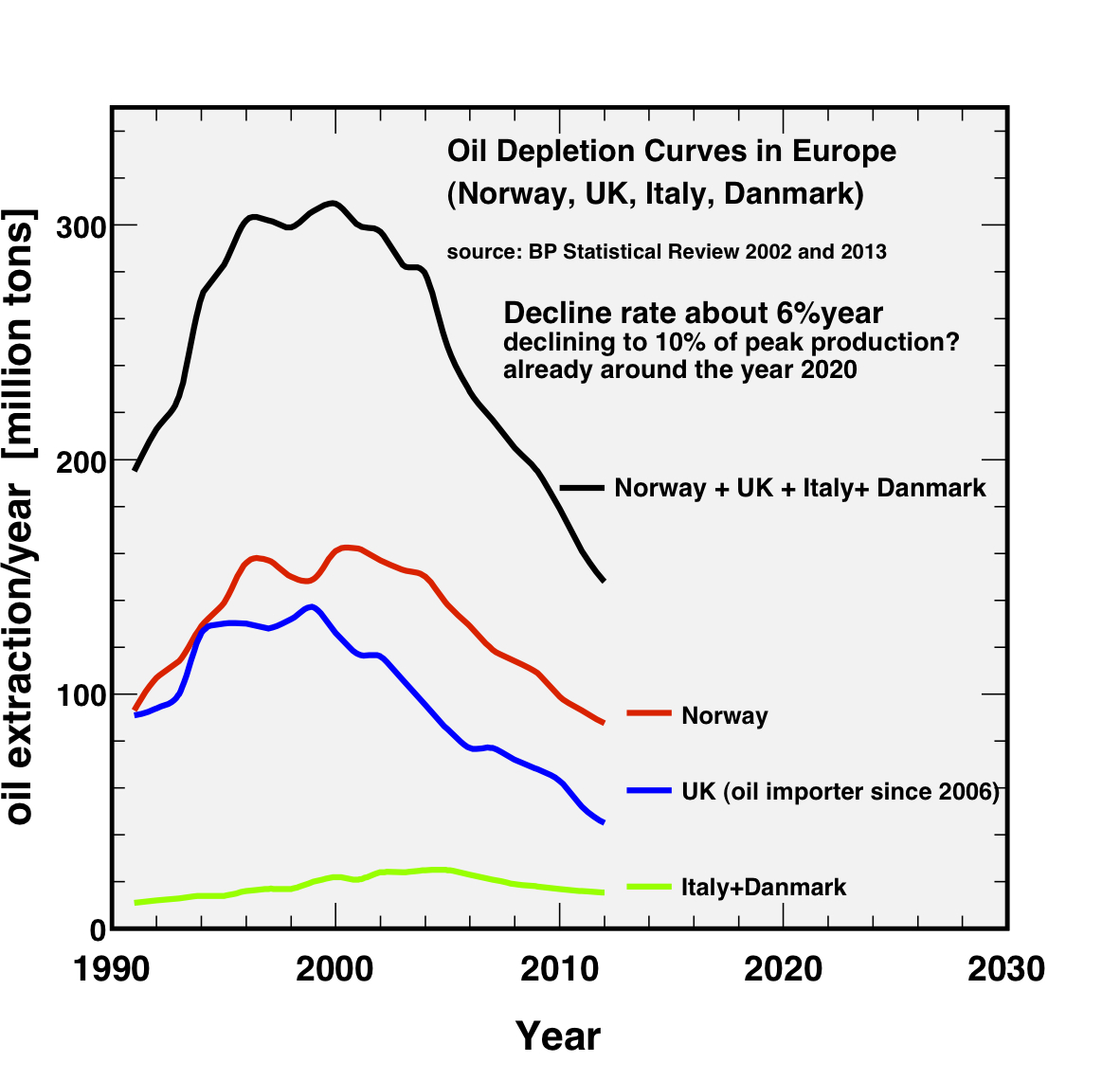}
\caption{Oil extraction curves from different larger oil rich countries and regions in Western Europe. 
Steep decline rates, about 5\% annually, are observed despite the factor of four higher oil prices since the 
year 2003, and the fact that the domestic sources are supplying only about one quarter of the Western European oil demand.}  
\end{center}
\end{figure}
When one considers (1) the current importance of oil for the entire way of life in Switzerland and its 
surrounding countries; (2) that the oil extraction from the Western European oil fields is declining 
by roughly 5\% every year, \cite{BPreport}, (see Figure 2), despite the fact that the price for a barrel of oil has increased 
by a factor of four; (3) that it is totally unknown how the oil based transport system 
can be modified, without loosing dramatic proportions of its services; and (4) that the oil 
dependence of the Western European way of life seems to be the most important unsustainable achilles heel, it becomes clear that a ``development 
towards sustainability" roadmap must concentrate on policies and actions that reduces the oil 
dependance of Switzerland and Western Europe, if disasters are to be avoided.      

Another important and rarely discussed unsustainable key area for Switzerland is its ``food energy" supply 
system, largely based on imports and its oil based production and distribution infrastructure. 
Its annual food import dependence has grown to more than 50\%, \cite{SwissFood}, and is 
certainly much higher during the winter months. Any instabilities within the Western European transport sector, for example due to higher oil prices, might lead to higher food prices and even supply disruptions.
It might thus be a good idea to prepare for the possibility that the current industrialized agricultural food oversupply situation might turn quickly into dramatic shortages within Switzerland and other regions in Western Europe.   

Concerning the negative impact from our industrial agriculture system, 
some alternative agricultural methods, based on sustainable
local production and distribution systems, have been developed during the past 
decades \cite{permaculture}. Successful demonstration projects in many regions and climates of our planet
have shown that it is theoretically possible to transform an agricultural system within a few years 
to produce better food and timber and to do so with yields per m$^{2}$ that are comparable to those of the present industrial agricultural system. In addition, when applied carefully,
the same methods allow us to turn unproductive and heavily eroded lands back into well 
functioning and productive systems that benefit the entire ecosystem as well as our species.

Such methods can be considered as some kind of potential ``interest rate", for the 
renewable part of the natural capital only. Such contributions must be added to our 
concept which relates the time dependent human impact to the remaining natural capital according to  
formula 2 below.  We thus observe that the time (t) dependent natural capital can be  
reduced by destructive type 1 human impacts ($I_{1}(t)= P_{1}A_{1}T_{1}$),
and increased in principle by ``regeneration" type 2 impacts ($I_{2}(t)= P_{2}A_{2}T_{2}$).   
Type 2 impacts, which increase natural capital growth rates, depend thus also on the number of local, regional and global people contributing actively to productive repair mechanisms. 

\begin{center}
\begin{equation}
{renewable~capital(t) = renewable~capital (t=0) - I_{1}(t) + I_{2}(t)} 
\end{equation} 
\end{center}
~\\
It thus turns out, then, that following our ``development towards sustainability" roadmap, we have encountered a lucky situation in which some human impacts not only reduce the current 
momentum towards disaster but enable us to cooperate with the renewable part of the natural
capital and to do so in such a way that the natural capital is able to satisfy our local needs 
while simultaneously enriching itself. 
We can thus conclude this section with the idea that ``development towards sustainability" roadmaps 
must lead to a reduction of all unsustainable practices, type 1 human impacts, and this with a 
certain yet to be determined minimal speed. In addition, such a roadmap must encourage the growth of all type 2 human impacts. Indeed, considering the urgency to act, we should launch a program even larger than the ÒManhattan ProjectÓ, a program that enables us to research the scientiÞc basis for type 2 human impacts and at the same time enables us to disseminate what is learned to educate people everywhere about this positive sort of impact.

\section{Summary}

The human impact on the environment, as deÞned by the IPAT equation proposed in 1971 by Holdren and Ehrlich, has grown dramatically, especially during the last 50-100 years.
As a result, the original renewable and non-renewable natural capital, our resource base, has been reduced to such an extent that humanity is now threatening the functioning of the entire biosphere. 
The observable results are mass extinctions of species and the collapse of land- 
and sea-based ecosystems that are the basis for our own existence. So few within the scientiÞc 
community doubt that what we are doing is unsustainable that they are well described as 
``statistical outliers".
	
Vague policies addressing ``sustainable development" have been adopted by most 
governments on the planet in the two decades since the first Rio earth summit. But no matter 
what intentions were behind such policies, the quantiÞably worsening state of the planet's
non-renewable and potentially renewable natural capital, as revealed at the 2012 RIO+20 
conference, shows beyond doubt that we have failed to do what needed to be done.

Assuming that the survival instincts of a well informed humanity will end our current 
fatalistic acceptance of our unsustainable way of life, we believe that alternative strategies
must be based on unambiguous deÞnitions of ``sustainability" and ``development towards 
sustainability". Such deÞnitions can be based on the well founded negation method.
Following this new approach, ``unsustainable" key areas of our way of life can be easily 
identiÞed. Furthermore, this approach allows us (1) to quantify how well new policies 
might move us into less unsustainable practices in every key area, and (2) to measure,
at regular intervals, how well the policies we adopt are actually doing at all levels of 
our global society.
 
Finally, the separation of our human impact into one which separates its effects 
on the remaining and original renewable and non renewable natural capital requires that  
we not only find a ``development towards sustainability" roadmap, but that we need to 
identify the minimal speed requirements for this transition in each of the unsustainable key areas. 
Our analysis demonstrates that in addition to the natural capital ``interest" rate, that 
the human impact can not only reduce the natural capital far beyond the ``interest" rate, 
but that a carefully designed and managed human impact can also increase the 
natural yields and the overall natural ``interest" and repair rates in previously damaged areas. 

Developing such a potentially successful roadmap will require a huge scientific project, 
a project even larger than the Manhattan Project, to explore the unknowns of the situation
regarding our natural capital, to take regional particularities into account, and thereby to
determine the needed transition speed requirements both globally and locally.
   
As it seems obvious to outside ``martian like" observers, all data are demonstrating that:
\begin{enumerate}
\item  The continuation of current policies, regardless of the used names, leads us
with increasing speed into a self-made global chaos and collapse;  
\item at least theoretically, we can change the direction, reduce our unsustainable practices 
in all key areas and even develop strategies that not only repair some of the already damaged natural 
capital, but simultaneously provide as with increasing yields fulfilling our needs on local and regional 
levels; 
\item neither the remaining unknowns about the actual status of our natural capital, nor inactions 
in surrounding countries and regions provide excuses not to begin immediately with the application of some of the easiest to implement most controllable local ``Development towards sustainability" roadmap elements. 
\end{enumerate}
In summary and acknowledging that the probability to apply a successful application of the  
``Development towards sustainability" roadmap  on a global level is very small, 
it can be stated, and beyond doubt, that our current policies and practices are leading,  
absolutely inevitably, to total collapse and to the extinction of countless species, our own quite possibly included.\\

\noindent

{\bf \large Acknowledgments\\} 

{\normalsize  \it{While the ideas presented in this paper are from the author alone, my thanks go to 
my friends, colleagues, and many students who have helped me during extensive exchanges to 
formulate these points and arguments in the current version. Especially I would like to thank W. Tamblyn and 
P. Martinez Ruiz del Arbol for their careful reading of the manuscript and for their suggestions 
on how to improve the clarity of the presented ideas. 
}
}
\newpage

\end{document}

As has been described in the introduction, ``Sustainability" and ``Sustainable development"  became widely used words during the last 20-30 years. Unfortunately, the original ideas got quickly lost, when used outside their original scientific context. In a similar way, theoretical interesting concepts like: (a) the overall regional and global human impact quantified with the I=PAT equation; (b) the carrying capacity with the possible temporary local and global overshoot and the resulting die-off problem; as well as (c) the idea to measure the ``ecological footprint" with respect to an abstract ``biological regeneration capacity", lost their original meaning once used outside of their original scientific environment.